# Quantum Spin-quantum Anomalous Hall Effect with Tunable Edge States in Sb Monolayer-based Heterostructures

*Tong Zhou[1], Jiayong Zhang[1], Yang Xue[1], Bao Zhao[1], Huisheng Zhang[1],*

*Hua Jiang[2,\*], and Zhongqin Yang[1,3,\*]*

1 State Key Laboratory of Surface Physics and Key Laboratory for Computational Physical Sciences (MOE) & Department of Physics, Fudan University, Shanghai 200433, China

2 College of Physics, Optoelectronics and Energy Soochow University, Suzhou (215006), China

3 Collaborative Innovation Center of Advanced Microstructures, Fudan University, Shanghai 200433, China

*Address correspondence to: jianghuaphy@suda.edu.cn, zyang@fudan.edu.cn

**ABSTRACT:**

A novel topological insulator with tunable edge states, called quantum spin-quantum anomalous Hall (QSQAH) insulator, is predicted in a heterostructure of a hydrogenated Sb ($Sb_2H$) monolayer on a $LaFeO_3$ substrate by using *ab initio* methods. The substrate induces a drastic staggered exchange field in the $Sb_2H$ film, which plays an important role to generate the QSQAH effect. A topologically nontrivial band gap (up to 35 meV) is opened by Rashba spin-orbit coupling, which can be enlarged by strain and electric field. To understand the underlying physical mechanism of the QSQAH effect, a tight-binding model based on $p_x$ and $p_y$ orbitals is constructed. With the model, the exotic behaviors of the edge states in the heterostructure are investigated. Dissipationless chiral charge edge states related to one valley are found to emerge along the both sides of the sample, while low-dissipation spin edge states related to the other valley flow only along one side of the sample. These edge states can be tuned flexibly by polarization-sensitive photoluminescence controls and/or chemical edge modifications. Such flexible manipulations of the charge, spin, and valley degrees of freedom provide a promising route towards applications in electronics, spintronics, and valleytronics.

# I. INTRODUCTION

The quantum anomalous Hall (QAH) effect [1-3] and quantum spin Hall (QSH) effect [4-6] have attracted considerable attention in condensed matter physics and material science, due to their dissipationless charge and spin current at the sample edges, respectively. Greatly potential applications are expected in electronics and spintronics based on the two effects [7]. Besides the charge and spin degrees of freedom, the new valley degree of freedom of electrons in honeycomb lattices has been recently predicted, which provides us another way to control electrons, known as valleytronics [8,9]. The (quantum) valley Hall ((Q)VH) effect [10-12], where Dirac fermions in different valleys flow to opposite transverse edges when an in-plane electric field is applied [10], paves the way to electric generation and detection of valley polarization in valleytronics. The QAH and QSH effects actually manifest the existence of the nontrivial topological states in the system, identified by Chern number (C) [13] and $Z_2$ [4], respectively. For QSH effect, especially in the case without the protection of the time reversal symmetry (TRS), spin Chern number ($C_S$) can be employed to describe the effect [14]. It is, however, not as fundamental as $Z_2$ and is a true topological invariant only after an extra criterion is satisfied. Namely, the spin spectrum gap should remain gapped [14, 15]. The QVH effect can be described by valley Chern number ($C_V$), defined as $C_V = C_K - C_{K'}$ [12], which gives the different distributions of the Berry curvatures around the two valleys (K and K'). Generally speaking, only one of the QAH, QSH, and QVH effects can be realized in a specific system. It would be thus very interesting if these three effects can be achieved in one single system, where the degrees of freedom of the carriers can be chosen as charge, spin, or valley as desired.

These interesting behaviors may be realized based on quantum spin-quantum anomalous Hall (QSQAH) effects, where the QAH state is expected to occur at one valley and the QSH state occurs at the other valley, proposed first by Ezawa with tight-binding (TB) models [16]. It is, however, full of challenges to realize the effect based on Ref. 16 due to some theoretical assumptions made and weak intrinsic spin-orbit coupling (SOC) in the proposed silicene system. Recently, a fully hydrogenated Sb (SbH) monolayer has been proposed to be a two dimensional (2D) QSH insulator with a pretty large band gap by Song et al, which can be stable even at 400 K obtained from molecule dynamics simulations [17]. The relatively large intrinsic SOC of Sb $p_x$ and $p_y$ orbitals in the system [18] should be beneficial to lead to the interesting topological effect. To produce the QSQAH effect, a stagger exchange field is indispensable [16], which may be introduced through adsorbing transition metal (TM) atoms [19]. The magnetic atoms adsorbed on film surfaces, however, usually tend to form into clusters [20,21], and as a result the stagger exchange field may not survive as expected. Up to the present, the behaviors of the edge states of the QSQAH effect have not been investigated, which are as significant as the studies of the topological invariants for topological materials since it is the edge states that conduct the charge, spin, or valley currents in the devices.

In this work, we tend to achieve the interesting QSQAH state and explore the behaviors of its edge states in a hydrogenated Sb monolayer epitaxial growth on a LaFeO$_3$ substrate (Sb$_2$H/LaFeO$_3$). LaFeO$_3$ is a G-type antiferromagnetic (AFM) insulator with the Fe sites forming alternating (111) ferromagnetic planes [22,23]. The LaFeO$_3$ (111) film has been experimentally synthesized with very high quality [24], whose lattice matches well the hydrogenated Sb film lattice with the mismatch of ~4% [17,23,24]. Through the proximity effect [25,26], the LaFeO$_3$ substrate is found to magnetize well the Sb$_2$H monolayer and further induce a staggered exchange field in the Sb$_2$H monolayer. Based on density-functional theory (DFT) and Wannier function methods, we demonstrate an exotic QSQAH effect in the Sb$_2$H/LaFeO$_3$ heterostructure. To understand the mechanisms, a tight-binding model based on $p_x$ and $p_y$ orbitals is constructed for the first time. From the built model, the edge states of the QSQAH effect are studied systematically. The results reveal that the dissipationless chiral charge current related to one valley emerges along the both sides of the sample, while the low-dissipation spin current related to the other valley is flowing only along one of the sample sides, a very unique phenomenon for the spin current in the effect. A QSQAH device prototype is also proposed from the heterostructure in which the edge states can be tuned flexibly by chemical edge modifications and/or polarization-sensitive photoluminescence controls [27-29]. Our work provides the deep understanding of the new mechanism of the QSQAH effect triggered by $p_x$ and $p_y$ orbitals and the exotic edge states of the effect.

## II. COMPUTATIONAL METHODS

The geometry optimization and electronic structure calculations were performed by using the first-principles method based on density-functional theory (DFT) with the projector-augmented-wave (PAW) formalism [30], as implemented in the Vienna ab-initio simulation package (VASP) [31]. The Perdew-Burke-Ernzerhof generalized-gradient approximation (GGA) was used to describe the exchange and correlation functional [32]. The substrate with different thicknesses containing four to seven Fe layers is chosen (there are seven Fe layers in Fig. 1a). It was found that they gave the same results. To stabilize the substrate, the Fe terminal surface of the $LaFeO_3$ film, opposite to the $Sb_2H$ interface, was passivated by two fluorine atoms per unit cell. A vacuum space of larger than 15 Å was used to avoid the interaction between two adjacent heterostructure slabs. For the structural relaxation, the $Sb_2H$ sheet and the topmost Fe, La, and O atoms in the substrate were allowed to relax until the Hellmann-Feynman force on each atom was smaller than 0.01 eV/Å. All calculations were carried out with a plane-wave cutoff energy of 550 eV and $12 \times 12 \times 1$ Monkhorst-Pack grids were adopted for the first Brillouin zone integral. To take into account the correlation effects of Fe $3d$ electrons, the GGA+$U$ method was adopted and the value of the Hubbard $U$ was chosen to be 3.5 eV, which can describe the correlation effect of Fe $3d$ electrons in $LaFeO_3$ bulk material correctly. The magnetization axis of the FM state in the heterostructure is set as the $z$ direction (out of plane), since it is found to be more favored than that lying in the $xy$ plane by 3.1 meV. The van der Waals (vdW) interaction correction with Grimme (DFT-D2) method [33] was considered in the calculations, including the structural relaxation.

## III. RESULTS AND DISCUSSION

### A. Geometry and electronic structures

Fig. 1a illustrates the designed heterostructure, consisting of one hydrogenated Sb monolayer on a (111) surface of a $LaFeO_3$ thin film with seven Fe layers. The Fe atoms in each (111) plane of the $LaFeO_3$ film are ferromagnetic (FM), while the neighboring Fe layers have opposite spin directions.

Through the proximity effect, the Sb sheet can be magnetized by the nearby Fe layers in the substrate. As shown in Fig. 1a, half of the Sb ($Sb_A$) atoms in the Sb sheet connect to the topmost Fe atoms in the substrate and the other half of the Sb ($Sb_B$) atoms bond with H atoms. The large calculated adsorption energy (2.8 eV) for the configuration indicates a very strong interaction between the $Sb_2H$ monolayers and the substrate. This tendency is consistent with the medium average distance $d_0$ (2.66 Å) between the $Sb_2H$ monolayers and the topmost layer of Fe atoms in the substrate after the geometry optimization.

Due to this strong interaction, a very distinct exchange field is generated in the Dirac bands of the $Sb_2H$ monolayer, and the bands around the K and K' points near the $E_F$ are spin polarized drastically, leading to inversion of the bands around the Dirac cones (Fig. 2a). It is also interesting to find that the Dirac bands of the hydrogenated Sb monolayer [17] are just located inside the bulk band gap (2.1 eV) of the $LaFeO_3$ film [34], making the Dirac bands primarily determine the electronic transport of the heterostructure. After the SOC is considered, local energy gaps of about 22 and 268 meV are opened around the K and K' points, respectively (Fig. 2b). The different features occurring at the K and K' points originate from the inequality of the $Sb_A$ and $Sb_B$ atoms which connect with magnetic Fe and the non-magnetic H atoms, respectively (Fig. 1). It seems that the QAH effect happens at K point and the QSH effect occurs at K' point, meaning the QSQAH state [16] may emerge in the $Sb_2H/LaFeO_3$ heterostructure.

To deeply understand the magnetic origins of the Dirac bands, the partial densities of states (PDOSs) of the Sb and the topmost Fe atoms are displayed in Fig. 2c,d, respectively. With the hydrogenated Sb monolayer adsorbed on the (111) surface of $LaFeO_3$ substrate, the $p_z$ states of $Sb_A$ atoms are strongly spin-polarized, while the $p_z$ states of $Sb_B$ atoms are almost spin-degenerate (Fig. 2c), indicating the exchange field at $Sb_A$ sites is much larger than that at $Sb_B$ sites. This tendency is ascribed to $Sb_A$ bonding with magnetic Fe ions and $Sb_B$ bonding with non-magnetic H ions (Fig. 1a). The strong interactions between the $Sb_A$ $p_z$ and Fe $d_{z2}$ states can also be illustrated by their PDOS distribution in the energy region from −2 eV to 1 eV in Fig. 2c,d. Therefore, a drastic staggered exchange field is triggered in the functionalized Sb monolayer, resulting in the sharp different band features near the $E_F$ occurring at the K and K' points in Fig. 2b, one of the most significant factors to produce the QSQAH state. The analysis of the bands and PDOSs also gives that the bands around the Dirac cones are mostly composed of degenerate Sb $p_x$ and $p_y$ states, which can lead to stronger intrinsic SOC interactions than the usual $p_z$ orbitals do in graphene and silicene etc [12,25].

**B. Topological identifications**

Topological invariants of Chern number and spin Chern number $C_S$ are calculated to identify whether the bands of Sb$_2$H/LaFeO$_3$ heterostructure (Fig. 2b) emerge the QSQAH effect. In our calculations, C and Cs are calculated by $C = \frac{1}{2\pi}\sum_n \int_{BZ} d^2k \Omega_n$ [7,35] and $C_S = \frac{1}{2\pi}\sum_n \int_{BZ} d^2k \Omega_n^S$ [36,37], respectively. The Berry curvature $\Omega(\mathbf{k})$ is calculated by [13,35]

$$\Omega(\mathbf{k}) = \sum_n f_n \Omega_n(\mathbf{k}),$$
$$\Omega_n(\mathbf{k}) = -2\,\text{Im} \sum_{m \neq n} \frac{<\psi_{n\mathbf{k}}|v_x|\psi_{m\mathbf{k}}><\psi_{m\mathbf{k}}|v_y|\psi_{n\mathbf{k}}>\hbar^2}{(E_m - E_n)^2}. \quad (1)$$

The spin Berry curvature $\Omega^s(\mathbf{k})$ is obtained by [36,37],

$$\Omega^s(\mathbf{k}) = \sum_n f_n \Omega_n^s(\mathbf{k}),$$
$$\Omega_n^s(\mathbf{k}) = -2\,\text{Im} \sum_{m \neq n} \frac{<\psi_{n\mathbf{k}}|j_x^s|\psi_{m\mathbf{k}}><\psi_{m\mathbf{k}}|v_y|\psi_{n\mathbf{k}}>\hbar}{(E_m - E_n)^2}. \quad (2)$$

In Eqs. (1) and (2), $E_n$ is the eigenvalue of the Bloch functions $|\psi_{n\mathbf{k}}>$, $f_n$ is the Fermi-Dirac distribution function at zero temperature, and $j_x^s$ is the spin current operator defined as $(s_z v_x + v_x s_z)/2$, where $v_x$ and $v_y$ are the velocity operators and $s_z$ is the spin operator. Note that the spin Berry curvature defined here has same unit as that of the Berry curvature, both of which are calculated in Wannier function bases [38] with the maximally localized algorithm [39].

The obtained Berry curvature $\Omega(\mathbf{k})$ and spin Berry curvature $\Omega^s(\mathbf{k})$ along the high-symmetry lines and their distributions in the 2D momentum space are plotted in Fig. 3. The peaks of the Berry curvatures $\Omega(\mathbf{k})$ are only primarily located at the K point with the band inversion (Fig. 3a). The corresponding distribution of $\Omega(\mathbf{k})$ in the momentum space is given in Fig. 3b, indicating an imbalanced distribution of Ω(k) between the valleys K and K' in the first Brillouin zone (BZ). After integration of the Berry curvatures throughout the whole first BZ as well as around each individual valley, we obtain the Chern number C = 1 as well as $C_K$ = 1 and $C_{K'}$ = 0, demonstrating its nontrivial topological features of valley polarized QAH phases ($C_V$ = $C_K$ − $C_{K'}$ = 1) [12]. In Fig. 3c,d, the spin Berry curvature $Ω^s$(**k**) only distributes around the K' point with no band inversion (The integral of $Ω^s$(**k**) around the K point is counteracted due to the both positive and negative values of $Ω^s$(**k**) around this point). By integrating $Ω^s$(**k**) over the BZ, spin Chern number $C_s$ ~ 1/2 with $C_{s,K}$ = 0 and $C_{s,K'}$ ~ 1/2 is obtained. The $C_s$ ~ 1/2 means the QSH effect is valley-polarized, only occurring at K' valley. For QSQAH systems, Cs should be exactly 1/2. However, due to the Rashba SOC, the spin is no longer a good quantum number in the SbH/LaFeO$_3$ heterostructure, resulting in the calculated $C_S$ not exactly equal to 1/2. Since the Rashba SOC is much less than the intrinsic SOC (~8%, see the caption of Fig. 4c) in our work, the approximation is acceptable. Note that when the proper spin sectors are chosen and the Cs is calculated as $C_S$= 1/2($C_↑$ − $C_↓$) [40], the obtained Cs would be exactly 1/2. These results prove that the Sb$_2$H/LaFeO$_3$ heterostructure is indeed a QSQAH insulator: the QAH effect occurring at the K point while the QSH effect happening at the K' point.

The influence of different surface decorations on the topological state of Sb monolayer-based heterostructure is also explored. The QSQAH state still remains robust when the $p_z$ orbitals of Sb$_A$ atoms are saturated with halogen atoms (Cl and Br). The bands are calculated and plotted in Fig. S1 of Supplemental Material [41], where the bands around the $E_F$ are very similar to those of Sb$_2$H/LaFeO$_3$. These results provide flexibility to fabricate the heterostructures and observe the QSQAH effect in experiments.

**C. Tight-binding models**

To understand well the QSQAH effect obtained in the heterostructure, a TB model is constructed. The TB model of QSQAH effects proposed in Ref. 16 is based on the $p_z$ orbitals, while the QSQAH state achieved in the Sb$_2$H/LaFeO$_3$ heterostructure is related to $p_x$ and $p_y$ orbitals. Generally, the physics effect based on $p_x$ and $p_y$ orbitals are much different from that of $p_z$ orbitals. The characteristic flat bands (see Fig. S2 of Supplemental Material [41]) and very large intrinsic SOC can appear in $p_x$ and $p_y$ systems [18,42]. Thus, the mechanism of the QSQAH effect in the Sb$_2$H/LaFeO$_3$ heterostructure may be different from that in Ref. 16. The spherical harmonic function $|\phi_+\rangle = -\frac{1}{\sqrt{2}}(p_x+ip_y)$ and $|\phi_-\rangle = \frac{1}{\sqrt{2}}(p_x-ip_y)$ orbitals are adopted as the basis here [43]. The TB Hamiltonian under the basis $\Phi_i = \{|\phi_+\rangle, |\phi_-\rangle\} \otimes \{\uparrow, \downarrow\}$ can be written as:

$$H = U\sum_i \mu_i \Phi_i^+ \tau_0 \otimes \sigma_0 \Phi_i + \lambda_{so} \sum_i \Phi_i^+ \tau_z \otimes \sigma_z \Phi_i + M_{A(B)} \sum_{i \in A(B)} \Phi_i^+ \tau_0 \otimes \sigma_z \Phi_i$$
$$+ (\sum_{i \in A} \sum_{\delta=1,2,3} \Phi_i^+ T_\delta \Phi_{i+\delta} + \sum_{i \in A} \sum_{\delta=1,2,3} \Phi_i^+ T_{R\delta} \Phi_{i+\delta} + H.C.) \quad (3)$$

Here, $\Phi_i$ represents annihilation operator on site i. Both $\tau_0$ and $\sigma_0$ are $2\times 2$ unitary matrices. $\tau_z$ and $\sigma_z$ indicate the Pauli matrices acting on orbital $\{|\phi_+\rangle, |\phi_-\rangle\}$ and spin $\{\uparrow, \downarrow\}$ spaces, respectively. $\delta_1 = (1,0)$, $\delta_2 = (-\frac{1}{2}, \frac{\sqrt{3}}{2})$, and $\delta_3 = (-\frac{1}{2}, -\frac{\sqrt{3}}{2})$ are the three vectors to the nearest neighbor sites. The first term in Eq. (3) represents the staggered potential with $\mu_i = 1$ ($-1$) for A (B) sublattice. The second term is the intrinsic SOC which is in on-site form. The third terms correspond to the magnetic exchange fields ($M_A$) and ($M_B$) at the Sb$_A$ and Sb$_B$ sites, induced by the LaFeO$_3$ substrate and the adsorbed H atoms, respectively. The fourth term represents the nearest-neighboring hopping. The last term represents the extrinsic Rashba SOC caused by the substrate. With the consideration of symmetry, matrices $T_\delta$ and $T_{R\delta}$ take the forms of

$$T_{\delta_1} = \begin{bmatrix} t_1 & t_2 \\ t_2 & t_1 \end{bmatrix} \otimes \sigma_0, \ T_{\delta_2} = \begin{bmatrix} t_1 & z^2 t_2 \\ z t_2 & t_1 \end{bmatrix} \otimes \sigma_0, \ T_{\delta_3} = \begin{bmatrix} t_1 & z t_2 \\ z^2 t_2 & t_1 \end{bmatrix} \otimes \sigma_0, \quad (4)$$

and

$$T_{R\delta_1} = -i\begin{bmatrix} \lambda_R & \lambda_R' \\ \lambda_R' & \lambda_R \end{bmatrix} \otimes \sigma_y, \ T_{R\delta_2} = i\begin{bmatrix} \lambda_R & z^2 \lambda_R' \\ z \lambda_R' & \lambda_R \end{bmatrix} \otimes (\frac{\sqrt{3}}{2}\sigma_x + \frac{1}{2}\sigma_y), \ T_{R\delta_3} = i\begin{bmatrix} \lambda_R & z \lambda_R' \\ z^2 \lambda_R' & \lambda_R \end{bmatrix} \otimes (-\frac{\sqrt{3}}{2}\sigma_x + \frac{1}{2}\sigma_y), \quad (5)$$

with $z = \exp(\frac{2}{3}i\pi)$. $t_1$, $t_2$ represent the hopping amplitudes, and $\lambda_R$, $\lambda_R'$ reflect the Rashba SOC between different orbitals of nearest-neighbor sites. In general, the relation of $\lambda_R$ and $\lambda_R'$ follows tendency of the $t_1$ and $t_2$. We assume $\lambda_R' = \frac{t_2}{t_1}\lambda_R$ for simplicity and to reduce the number of the independent parameters. Hence, there are totally seven independent parameters in the model. To our best knowledge, it is the first time to give the Rashba term in hexagonal lattices based on $p_x$ and $p_y$ orbitals.

**D. Competitive mechanism and tunable edge states**

The band evolution with the different parameters in the TB model is given in Fig. 4a-d and the corresponding behaviors of the edges states of the zigzag-edge nanoribbon are displayed in Fig. 4e-h. For the simplest case (Fig. 4a), since $M_A = M_B = 0$ and $\lambda_{SO} = 220$ meV, the system is a normal QSH insulator with the time reversal symmetry protected. There exist four edge states labeled as A, B, C, and D. A, C and B, D are both spin-degenerate (Fig. 4e). From the spatial distribution of the wave functions, one can find that states A and B are localized near one boundary of the ribbon, while C and D are localized near the other boundary. When a staggered exchange field ($M_A = 180$ meV and $M_B = 30$ meV) is applied to the system, three types of the topological phases will emerge. Namely, when $\lambda_{SO} >$ both $M_A$ and $M_B$ (Fig. 4b), two unequal QSH gaps are opened at K and K' point, respectively. A, B, C, and D edge states all still exist in the bulk gap, giving rise to a net spin transport (Fig. 4f). The system is in a QSH state ($C_S \sim 1$) with the time reversal symmetry broken [40,44], not as fundamental as QSH with TRS, due to the Cs definition. Note that there are small energy gaps appearing in the crossings of A, B and C, D due to the absence of TRS. For this case, an example can be carried out easily by replacing Sb film with Bi film to increase the $\lambda_{SO}$ in the system. As shown in Fig. S3 of Supplemental Material [41], a QSH state with TRS broken is achieved in the BiH/LaFeO$_3$ heterostructure. When $M_A > \lambda_{SO} > M_B$, a QAH gap is opened at the K point, while a QSH gap is opened at the K' point (Fig. 4c). Thus, the exotic QSQAH state is achieved (with C = 1 and Cs ~ 1/2). Edge states A and B still exist in the bulk band gap of the system, while the edge states C and D move outside of the bulk band gap, which will not contribute to the electronic transport of the system. Meanwhile, new chiral charge edge states E and F appear in the QAH gap (Fig. 4g). It is interesting to find that the edge states of E and F related to the K valley conduct the chiral charge current along the both sides of the sample, while the spin current (A and B) flows only along one side of the sample. The spin current flowing along the other side of the sample does not appear due to the edge states C and D outside of the electronic transport energy window. The sketch of the distribution of the edge states in this QSQAH effect can be seen in Fig. 5a. When $\lambda_{SO}$ < both $M_A$ and $M_B$, two QAH gaps are opened at K and K' point, respectively (Fig. 4d). Edge states A, B, C and D all move outside of the bulk band gap of the system, while another new pair of the edge states of G and H appear at the K' valley. Thus, two chiral conducting channels appear on each side of the sample in this case (Fig. 4h). The system is in a pure QAH state with Chern number of C = 2. The effects of the extrinsic Rashba SOC (the last term in Eq. (3)) on the band structure are also investigated. The red curves in Fig. 4c,d show the bands without the Rashba SOC. It is of significance to find that the QAH gap cannot be opened without the Rashba SOC interaction. Thus, the magnitudes of the QAH gaps are determined solely by the strength of the Rashba SOC. The TB parameters for the realistic $Sb_2H/LaFeO_3$ heterostructure are obtained by fitting the TB band structure to that acquired from the first-principles calculations near the Dirac points (Fig. 2b). The fitted bands are displayed in Fig. 4c. We also investigate the edge states of the armchair-edge nanoribbon (see Fig. S4 of Supplemental Material [41]). Dissimilar to the zigzag case, the bulk bands of the K and K' points fold to the Γ point in the armchair nanoribbon, namely, the valleys are not good quantum numbers now. Thus, the QSQAH effect cannot be observed directly by photoluminescence controls in experiments in the armchair nanoribbon despite the edge states still existing.

Since the C and D edge states are outside of the electronic transport energy window in the obtained QSQAH state (Fig. 4g), only the A and B edge states contribute to the spin current of the system, which are located only along the bottom side of the sample, as shown in Fig. 5a. The spin current along the upper side of the sample (C and D) disappears. Thus, the space symmetry of the spin current coming from the K' valley is broken, while it is not for the charge current coming from the K valley. To comprehend this intriguing phenomenon of the spin current, we vary the exchange fields of the outmost A and B sublattices (labeled as $M_{AE}$ and $M_{BE}$, respectively) in the nanoribbon (Fig. 5a) and plot the corresponding bands in Fig. 5b,c, respectively. Obviously, the charge edge states (E and F) are not affected at all by the $M_{AE}$ and $M_{BE}$. For the spin edge states, the $M_{AE}$ has no influence on the A and B edge states, located along the bottom edge of the nanoribbon, but affects significantly the C and D edge states because the C and D edge states also lie at the upper edge of the nanoribbon as the $M_{AE}$ does (Fig. 5b). The energy gap between the C and D edge states can be decreased drastically by decreasing the $M_{AE}$. When the $M_{AE}$ decreases from 180 meV to 30 meV, the energy gap between them disappears. Thus, the spin current flowing along the upper edge of the nanoribbon restores, conducting by the C and D edge states. Similarly, the $M_{BE}$ solely influences the spin edge states of A and B along the bottom edge of the nanoribbon (Fig. 5c). The energy gap between the A and B edge states changes, however, very slightly with the variation of $M_{BE}$, ascribing to the very large bulk band gap opened at the K' point. These results demonstrate that the space-symmetry-breaking of the spin edge state distribution is a unique characteristic of the QSQAH effect. It does not emerge in the pure QAH or QSH effects. The spin edge states usually only survive along the boundary with a weaker exchange field in the QSQAH system. The disappearing spin edge states, however, can restore by decreasing the exchange field along the other side of the sample. Therefore, the spin current in QSQAH insulators can be manipulated flexibly by such as chemical edge modifications [37, 45] in the system.

The results of Fig. 4 indicate that the competition of the $\lambda_{SO}$ and the $M_A$ ($M_B$) terms can control the topological phase transitions of the system. Since the extrinsic Rashba term is usually much smaller than the $M_A$ ($M_B$) and $\lambda_{SO}$ terms, this Rashba term can be neglected (set $\lambda_R = 0$) to better explore the competitive relationship between $M_A$ ($M_B$) and $\lambda_{SO}$. As provided in Supplemental Material [41], the topological phase transitions can be acquired by analyzing quantitatively the sequence of the energy levels around the $E_F$ at the K and K' points, which gives the same phase transition trend as obtained from Fig. 4. Thus, various topological phases can be achieved in the system by tuning the relative values of the $M_A$ ($M_B$) and $\lambda_{SO}$ terms. When $\lambda_{SO} = M_A$ ($M_B$), it gives the phase boundaries of the topological transitions.

The topological phase diagram with different magnitudes of the $\lambda_{SO}$, $M_A$, and $M_B$ is given in Fig. 6a. The QSQAH state obtained in Fig. 4c corresponds to the green region with $M_A > M_B$ (Cv = 1). If a system has $M_A < M_B$, the exotic QSQAH state still exists, but with the QSH effect at K point and QAH effect at K' point, namely Cv = −1, opposite to that in the case of $M_A > M_B$. One special status in Fig. 6a is that the QSQAH state does not emerge if $M_A = M_B$, indicating the indispensability of the staggered exchange field to produce the QSQAH effect. By comparing the dotted lines in Fig. 6a, it can be found that the more the staggered exchange field is, the easier the QSQAH effect is achieved. The staggered exchange field producing the QSQAH is somewhat similar to the valley Hall effect, induced by the local staggered electrostatic potential in the A and B sublattices, such as in graphene [46]. Hence, we conclude that this kind of the QSQAH state can be realized when a system owns the following elements: 1) a honeycomb lattice, 2) a staggered exchange field, 3) an intrinsic SOC with a suitable strength, which should be larger than the exchange field at the one set of the sublattice sites while smaller than the exchange field at the other set of the sublattice sites, and 4) sizable Rashba interactions, which determine the value of the topologically nontrivial gap opened.

Generally, large band gaps of topological insulators are desired to observe the effect at higher temperature in experiments [3]. Since the nontrivial band gap of the QAH effect at K point (Fig. 3a) is related to the strength of Rashba interactions in the system, the band gap can be enlarged by increasing the Rashba interactions. We apply an externally electric field to the heterostructure, which has been proved to be an effective tactics to tune the Rashba interaction. When the electric field is applied along -z direction, more charges transfer from LaFeO$_3$ substrate to Sb$_2$H sheet. The asymmetry of the Sb $p_x$ and $p_y$ orbital distribution above and below the Sb sheet becomes prominent, resulting in the increase of the Rashba interaction. Therefore, the band gap is increased as the electric field (along -z direction) is enhanced, as shown in Fig. 6b. When the electric field is up to −0.2V/Å, the band gap is about 28 meV, corresponding to the room temperature. Based on the same mechanism, the Rashba interaction in the Sb sheet can also be enhanced by reducing the distance between the Sb sheet and the LaFeO$_3$ substrate. As expected, the topologically nontrivial band gap is enlarged with the decrease of the distance between the Sb sheet and the substrate (Fig. 6b). When the distance is reduced by 4% ($d_0 - d = 0.1$ Å), the band gap is more than 26 meV.

## IV. EXPERIMENTAL PERSPECTIVE AND APPLICATIONS

In experiments, the Sb$_2$H/LaFeO$_3$ heterostructure designed in this work can be fabricated in three steps. 1) Grow the LaFeO$_3$ (111) films with the topmost Fe surface on such as a SrTiO$_3$ substrate [47]. Based on the mature technologies including the pulsed laser deposition (PLD) method and reflection–high energy electron diffraction (RHEED) etc., the fabricated LaFeO$_3$ (111) films can be atomic-scale controlled and have very high quality [24]. 2) Deposit Sb sheet [48,49] to the topmost Fe surface of the LaFeO$_3$ (111) film by epitaxial growth methods. Since the $p_z$ orbitals of the Sb atoms in the Sb monolayer are dangling, half of the Sb atoms (A sites) tend to form bonds with the outmost Fe atoms in the substrate. Note that the lattice constant of LaFeO$_3$ (111) surface is about 5.4 Å, on which the Sb atoms are favorable to form a planar honeycomb structure [50,51] as shown in Fig. 1a. The calculated adsorption energy of this pristine Sb film on the substrate is about 1.8 eV, indicating the Sb film with the planar honeycomb structure can form well on the substrate. 3) Expose the Sb/LaFeO$_3$ heterostructure to hydrogen or halogen plasma. This method has been successfully applied to synthesize graphane (hydrogenated graphene) from graphene [52]. Since the left half Sb atoms (B sites) in the monolayer are unbonded, these hydrogen or halogen atoms are inclined to adsorb and bond with the Sb$_B$ atoms rather than Sb$_A$ atoms, which can be confirmed by the total energy of H atoms on Sb$_B$ with 1.2 eV lower than that of H atoms on Sb$_A$. After the H atoms are adsorbed onto the Sb film surface, the finally obtained Sb$_2$H/LaFeO$_3$ heterostructure becomes more stable, with the Sb$_2$H adsorption energy of 2.8 eV, larger than that of the pristine of Sb film (1.8 eV). The reason can be ascribed to the adsorbed H atoms tending to enlarge the Sb lattice[14] and releasing the lattice distortion energy of the Sb film on the substrate. Thus, it is promising to fabricate the heterostructure of Sb$_2$H/LaFeO$_3$ in experiments. Actually, this epitaxial growth method have been a powerful tool to create new materials with desirable features in physics and chemistry [53,54].

As a potential application of the QSQAH heterostructure, a device prototype is proposed in Fig. 6c,d. With the polarization-sensitive photoluminescence control [27-29], the edge states can be variously manipulated. When the heterostructure is excited by the left-handed light, the dissipationless chiral charge current related to the K valley can be observed in the both edges of the heterostructure (Fig. 6c). While it is exposed in the right-handed light (Fig. 6d), the low-dissipation spin current related to the K' valley appears, which only flows along the sample side with a weaker exchange field. Moreover, by reducing the exchange fields of the other side of the sample, the spin current may also emerge at that boundary. Such flexible manipulations of the charge, spin, and valley degrees of freedom of the edge states provide a promising route towards applications in electronics, spintronics, and valleytronics.

## V. CONCLUSIONS

We systematically investigated the electronic and topological properties of the hydrogenated Sb monolayer epitaxial growth on the (111) surface of an AFM insulator $LaFeO_3$ from first-principles calculations and tight-binding models. The exotic QSQAH state is found in the heterostructure with the band gap opened exactly around the $E_F$. A tight-binding model based on $p_x$ and $p_y$ orbitals is constructed for the first time to deeply understand the mechanisms. The larger the staggered exchange field is, the easier the QSQAH effect can be obtained. The study of the edge states in the heterostructure shows that the charge current related to one valley emerges along the both sides of the sample, while spin current related to the other valley is flowing only along the sample side with a weak exchange field. The spin current along the other side of the sample can, however, be restored by tuning the local exchange field at the sample edge. A device based on the QSQAH heterostructure is designed, in which the degrees of freedom of the carries can be manipulated flexibly.

## ACKNOWLEDGEMENTS

This work was supported by National Natural Science Foundation of China with Grant Nos. 11574051, 11504008, and 11374219, NBRP of China with Grant No.2014CB920901, Natural Science Foundation of Shanghai with Grant No. 14ZR1403400, and Fudan High-end Computing Center.14

**Figures and captions**

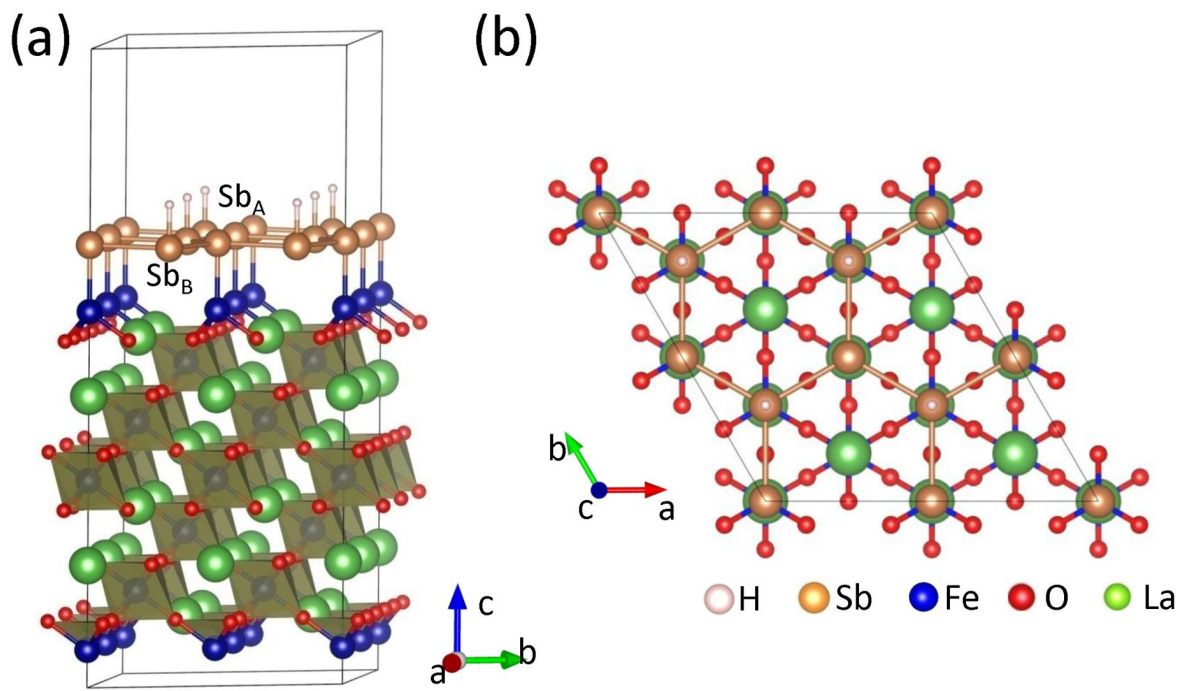

**Fig. 1**. (a) Side view of the Sb$_2$H/LaFeO$_3$ heterostructure. The H, Sb, Fe, O, and La atoms are denoted in pink, brown, blue, red, and green, respectively. (b) The top view of the heterostructure.

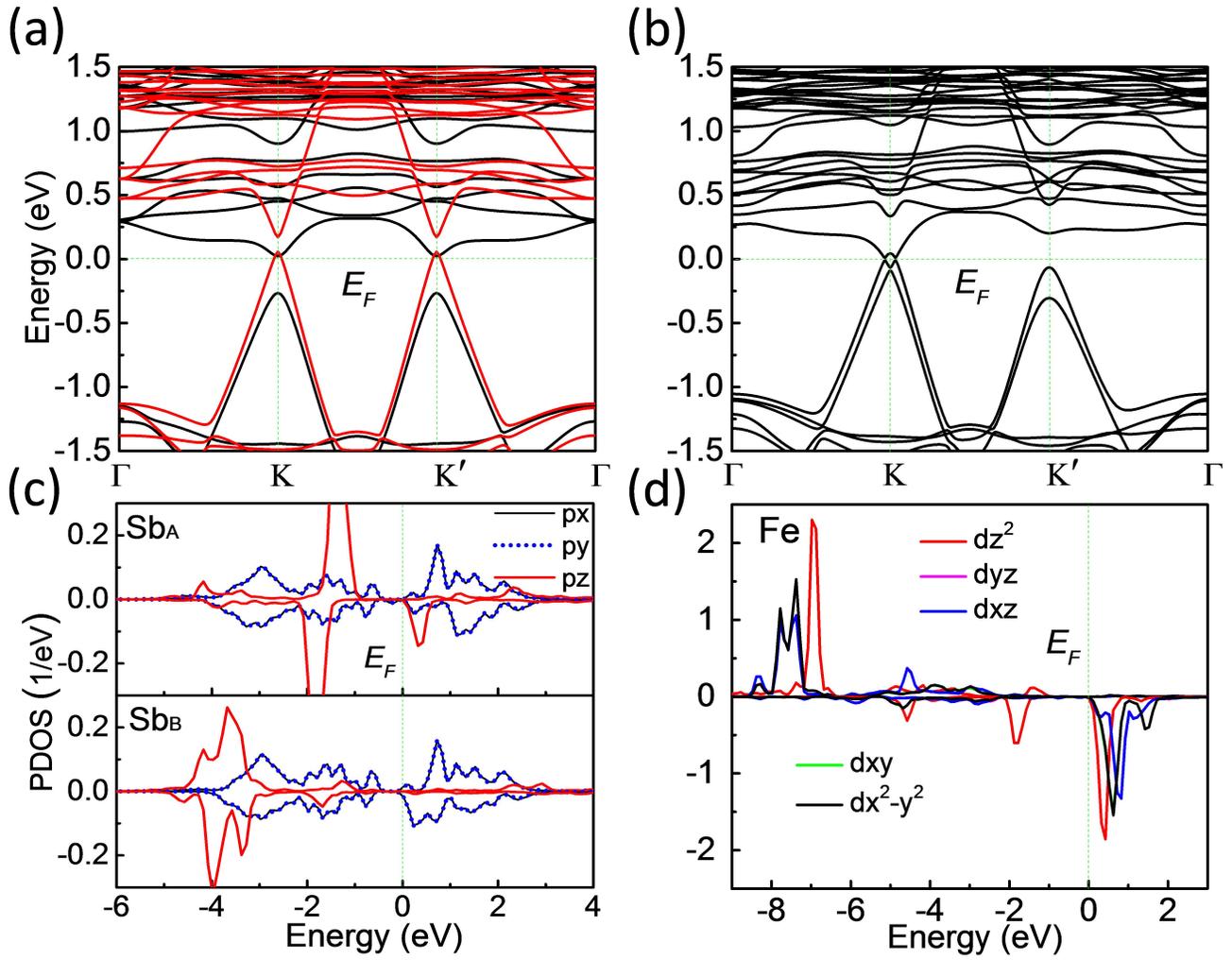

**Fig. 2**. (a) and (b) Band structures for the Sb$_2$H/LaFeO$_3$ heterostructure without and with SOC considered, respectively. The red and black curves in (a) denote the spin-up and spin-down states, respectively. (c) and (d) The calculated PDOSs of Sb and Fe atoms, respectively, without the SOC. The positive and negative values in (c) and (d) express the spin-up and spin-down states, respectively.

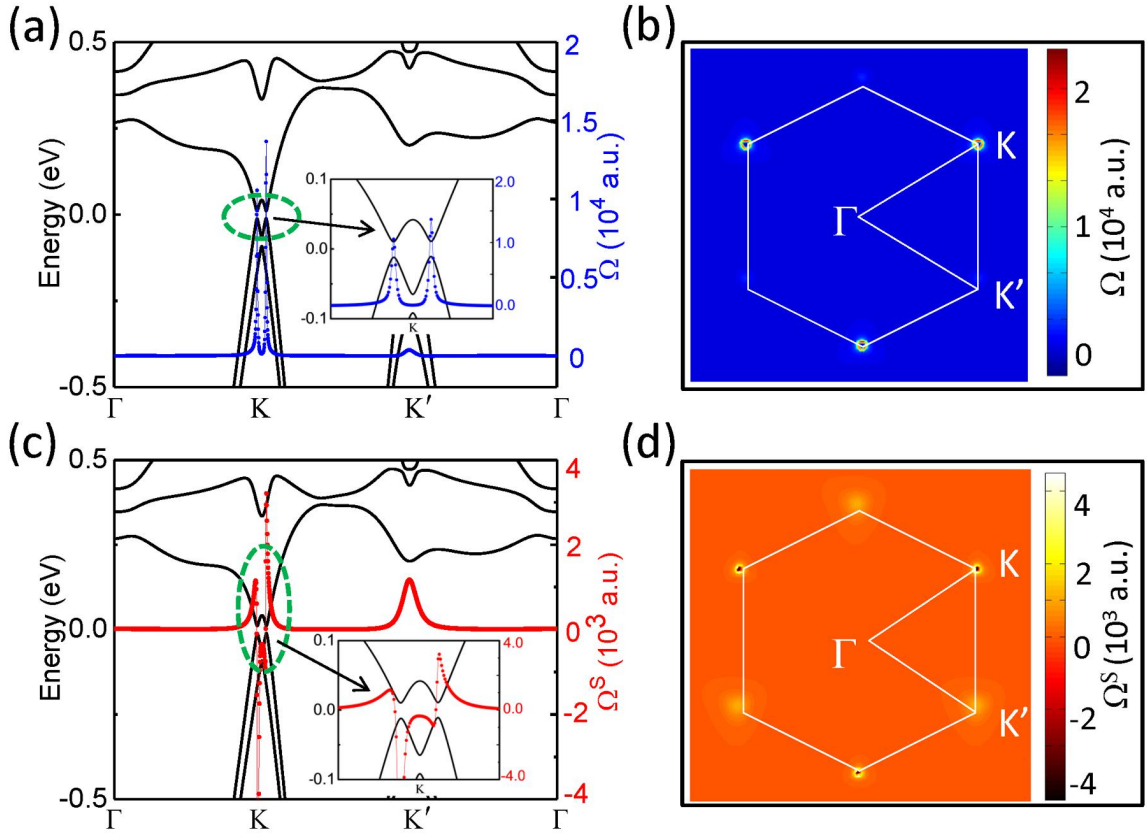

**Fig. 3**. (a) Band structures (solid black curves) for the $Sb_2H/LaFeO_3$ heterostructure. The blue dots denote the Berry curvatures (in atomic units (a.u.) for the whole valence bands. The inset in (a) is the magnified bands and Berry curvatures around the gap induced by the SOC interaction (b) The distribution of the Berry curvatures in 2D momentum space for the same system as in (a). (c) and (d) are the same as in (a) and (b), respectively, except for spin Berry curvatures (red dots in (c)) instead. The a.u. is related to SI unit as 1 a.u. = 0.28 Å$^2$.

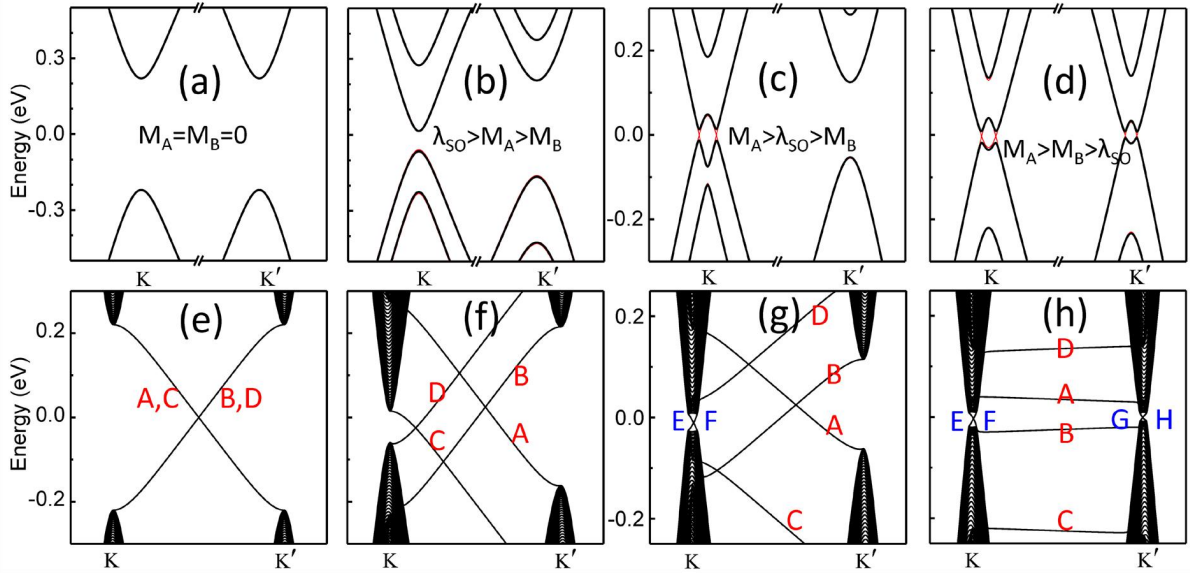

**Fig. 4**. (a) Band structure calculated from the TB model with $t_1 = -t_2 = 1$ eV, $U = M_A = M_B = \lambda_R = 0$, and $\lambda_{SO} = 220$ meV. (b) The same as (a) except for $U = -25$ meV, $M_A = 180$ meV, $M_B = 30$ meV, and $\lambda_R = 10$ meV. (c) and (d) The same as (b), except for $\lambda_{SO} = 120$ and 5 meV, respectively. The red curves in (b)-(d) denote the corresponding bands without the Rashba term ($\lambda_R = 0$). (e)-(f) The band structures of zigzag nanoribbons (containing 160 zigzag chains in the width) with the same TB parameters as for the black curves in (a)-(d), respectively. The TB parameters for the black curves in (c) are obtained by fitting the bands to the DFT results of Fig. 2b.

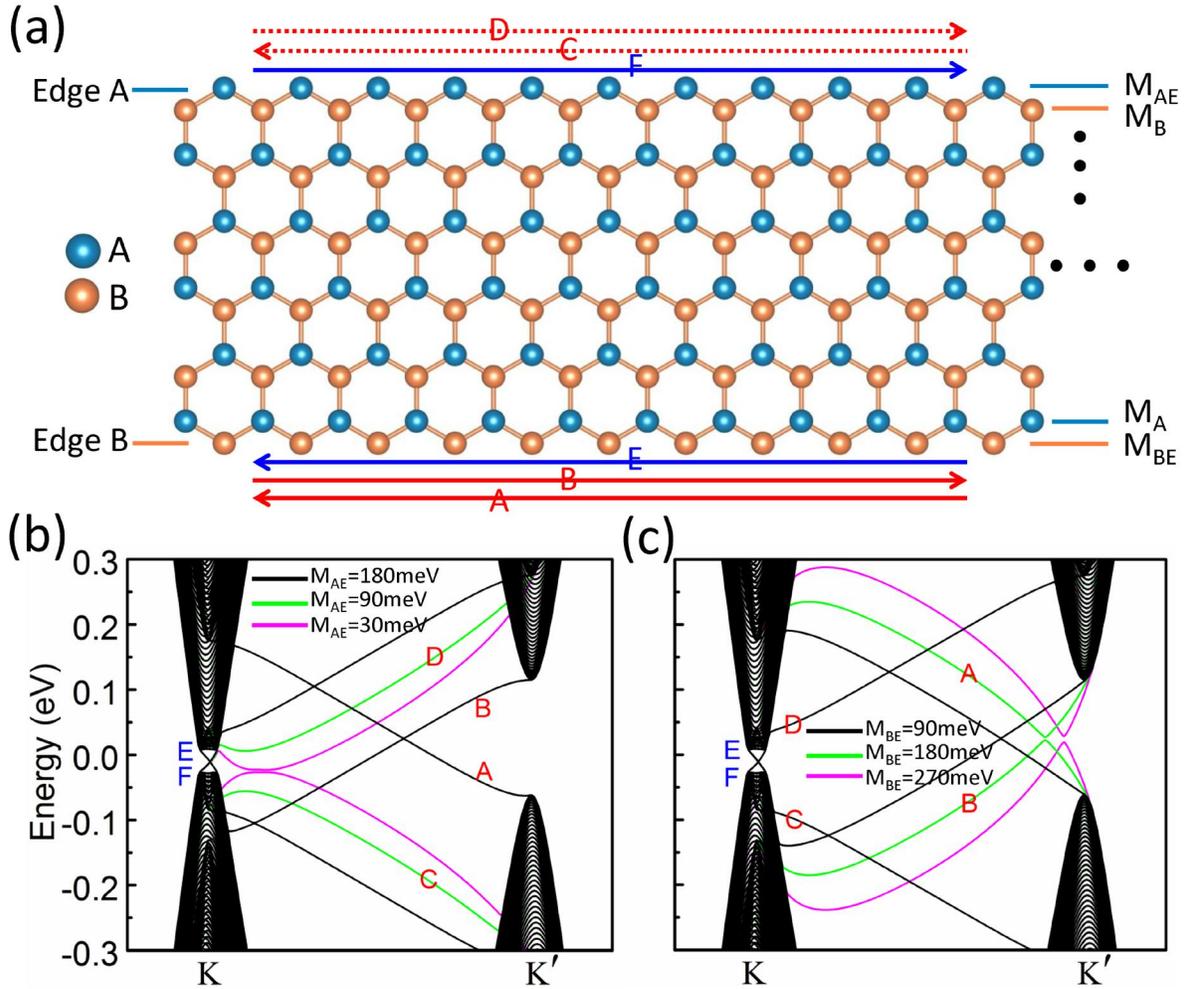

**Fig. 5**. (a) A schematic drawing of a zigzag nanoribbon (containing 160 zigzag chains in the width). Blue (brown) balls indicate the A (B) sublattice, of which the exchange field is $M_A$ ($M_B$). The exchange field of the outmost A (B) sublattice (Edge A (B)) is indicated by $M_{AE}$ ($M_{BE}$). The red and blue solid arrows indicate the spin and charge current flowing along the sample edges, respectively. The red dashed arrows indicate the disappearing spin current. (b) The band structures of the nanoribbon in (a) with the $M_{AE}$ decreasing from 180 meV to 30 meV, obtained from the TB model. The $M_{BE}$ is fixed at 30 meV. The other TB parameters are the same as those in Fig. 4c ($M_A$ = 180 meV and $M_B$ = 30 meV). (c) The same as (b) except for $M_{AE}$ = 180 meV and $M_{BE}$ = 90, 180, or 270 meV. A-F in (b) and (c) denote the edge states.

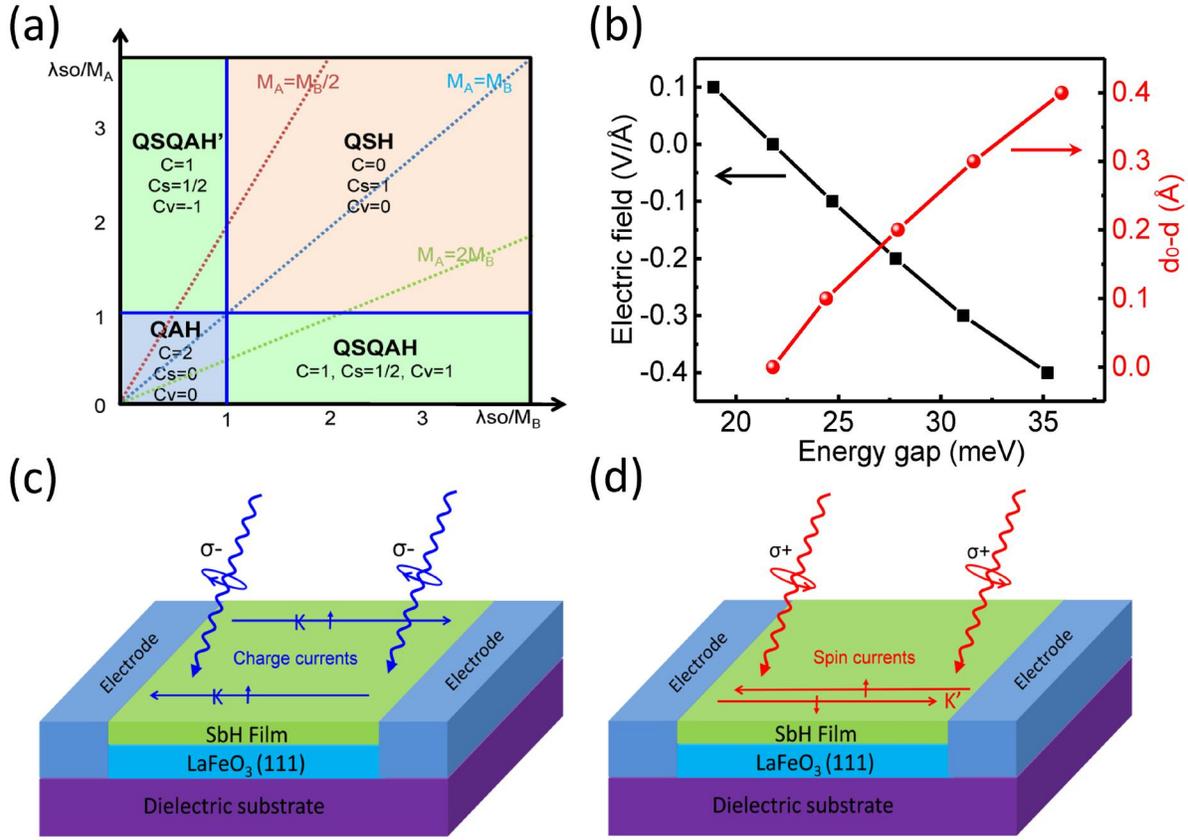

**Fig. 6**. (a) Phase diagram in the $\lambda_{SO}/M_A$-$\lambda_{SO}/M_B$ plane. Solid blue lines represent phase boundaries. There are three types of topological insulators: QAH (blue region), QSQAH (green region), and QSH (orange region). The orange, blue, and green dotted lines denote $M_A = M_B/2$, $M_A = M_B$, and $M_A = 2M_B$, respectively. (b) Black and red curves denote the band gaps as a function of the external electric field applied and the reduction of the separation between the Sb sheet and the LaFeO$_3$ ($d_0 - d$, where $d_0$ and $d$ is the distance between the Sb sheet and the substrate with and without the structural relaxation), respectively. (c) and (d) A schematic drawing depicting the QSQAH effect in the Sb$_2$H/LaFeO$_3$ heterostructure. With the left-handed (c) and right handed (d) light control, the charge (blue solid arrows) and spin edge states (red solid arrows) emerge along the edges of the heterostructure, respectively. The small solid arrows on the edge states indicate the spin directions.

23